\documentclass[aps,11pt,prd,longbibliography,notitlepage,nofootinbib]{revtex4-1}
\usepackage{tikz}
\usepackage{feynmp-auto}
\usepackage{amsmath}
\usepackage{todonotes}
\usepackage[utf8]{inputenc}
\usepackage[T1]{fontenc} 
\usepackage{array}
\usepackage{bbm}
\usepackage{soul}	
\usepackage{xcolor}

\usepackage{epsfig}
\usepackage{color}
\usepackage{slashed}
\usepackage{comment}
\usepackage{epstopdf}
\usepackage{xspace}
\usepackage{mathrsfs}
\usepackage[euler]{textgreek}
\usepackage{amsmath}
\usepackage{amsthm}
\usepackage{amsfonts}
\usepackage{amssymb}
\usepackage{graphicx}
\usepackage[font={small}]{caption}
\usepackage{subcaption}
\usepackage{float}
\usepackage{multirow}
\usepackage[hang, flushmargin,bottom]{footmisc} 
\usepackage{placeins}
\usepackage{enumitem}
\usepackage{slashed}
\usepackage{bbm}
\usepackage{appendix}
\usepackage{tabularx}
\usepackage{siunitx}

\usepackage{titlesec}
\titleformat{\chapter}[display]
  {\normalfont\LARGE\bfseries}
  {\chaptertitlename\ \thechapter}{5pt}{\LARGE}
  \titlespacing*{\chapter}{0pt}{-20pt}{35pt}
\usepackage{bigstrut}
\usepackage{pst-node}
\usepackage{pstricks}
\usepackage{physics}
\usepackage{mathtools}
\usepackage{setspace}
\usepackage{fancyhdr}
\usepackage[makeroom]{cancel}
\usepackage{feyn}
\newcommand{\be}{\begin{equation}}
\newcommand{\ee}{\end{equation}}
\newcommand{\bes}{\begin{equation*}}
\newcommand{\ees}{\end{equation*}}

\usepackage{hyperref}
\hypersetup{%
  colorlinks = true,
  linkcolor  = blue,
  citecolor  = blue,
  urlcolor   = blue,
}
\usepackage{soul}
\usepackage{xpatch}
\makeatletter
\xpretocmd{\todo}{\@bsphack}{}{}
\xapptocmd{\todo}{\@esphack}{}{}
\makeatother

\newcommand{\beq}{\begin{equation}}
\newcommand{\eeq}{\end{equation}}

\newcommand{\SU}{\,{\rm SU}}
\newcommand{\U}{\,{\rm U}}

\usepackage{tikz}

\DeclareRobustCommand{\swatch}[1]{\tikz[baseline=-0.6ex]\node[fill=#1,shape=rectangle,draw=black,thick,minimum width=5mm,rounded corners=0.5pt](){};}

\newcommand{\pT}{\ensuremath{p_\mathrm{T}}\xspace}

\newcommand{\MET}{\ensuremath{p_T^\mathrm{miss}}\xspace}

\newcommand{\herwig}{H\protect\scalebox{0.8}{ERWIG}\xspace}
\newcommand{\rivet}{R\protect\scalebox{0.8}{IVET}\xspace}

\newcommand{\contur}{\textsc{Contur}\xspace}

\newcommand{\madgraph}{\textsc{MadGraph5}\xspace}

\definecolor{green}{HTML}{008000}
\definecolor{goldenrod}{HTML}{DAA520}
\definecolor{magenta}{HTML}{FF00FF}
\definecolor{silver}{HTML}{C0C0C0}
\definecolor{indigo}{HTML}{4B0082}
\definecolor{skyblue}{HTML}{87CEEB}
\definecolor{darkgoldenrod}{HTML}{B8860B}
\definecolor{orange}{HTML}{FFA500}
\definecolor{yellow}{HTML}{FFFF00}
\definecolor{saddlebrown}{HTML}{8B4513}
\definecolor{blue}{HTML}{0000FF}
\definecolor{turquoise}{HTML}{40E0D0}
\definecolor{yellow}{HTML}{FFFF00}
\definecolor{white}{HTML}{FFFFFF}
\definecolor{whitesmoke}{HTML}{F5F5F5}
\definecolor{hotpink}{HTML}{FF69B4}

\newcommand{\myComment}[1]{}

\begin{document}
\title{\LARGE{Quark-Lepton Unification Signatures}}
\author{Jon Butterworth$^{1}$, Hridoy Debnath$^{2}$, Pavel Fileviez P{\'e}rez$^{2}$, Peng Wang$^{1}$}
\affiliation{
$^{1}$Department of Physics and Astronomy, University College London, Gower St., London, WC1E 6BT, UK \\
$^{2}$Physics Department and Center for Education and Research in Cosmology and Astrophysics (CERCA), Case Western Reserve University, Cleveland, OH 44106, USA}
\email{j.butterworth@ucl.ac.uk, hxd253@case.edu, pxf112@case.edu, peng.wang.22@ucl.ac.uk}
\date{\today}

\begin{abstract}
  We investigate the collider signatures of the minimal framework for quark-lepton unification at a scale not far from the
  electroweak symmetry breaking scale.
  This theory predicts a rich spectrum of new fields, including one vector leptoquark, two scalar leptoquarks, a color-octet scalar, and an additional Higgs doublet.
  Neutrino masses are generated via the inverse seesaw mechanism,
  facilitating viable matter unification at the low scale.
  We find that this theory predicts that in many cases the dominant leptoquark decays are to the third generation Standard Model quarks and leptons.
  We identify key experimental signatures at the Large Hadron Collider, evaluate and discuss the limits from
  current measurements, and outline potential strategies for probing this theory in the near future.
\end{abstract}

\maketitle

\section{INTRODUCTION}
The idea of quark-lepton unification~\cite{Pati:1974yy,Pati:1974yy,Pati:1973rp} is one of the most appealing ideas for physics beyond the
Standard Model (SM).
In this context, the SM quarks and leptons are unified in the same fermionic multiplets, and neutrinos are predicted to be massive.
This theory was in fact the first theory to be proposed for physics beyond the SM, also predicting for the first time that neutrinos
can be massive particles.
Unfortunately, the original theory predicts that the SM neutrinos and the up-quarks have the same mass, and the charged leptons and down
quarks have the same mass.
One can add a new Higgs field to solve the second problem, but since neutrino masses are much smaller than the other fermion masses in the SM,
one needs an extra mechanism for neutrinos to avoid extreme fine-tuning.
The so-called canonical seesaw mechanism~\cite{Minkowski:1977sc,Gell-Mann:1979vob,Mohapatra:1979ia,Yanagida:1979as} can be implemented in this framework, but since the Dirac neutrino mass term is predicted to equal 
the up-quark mass, the gauge symmetry for quark-lepton unification has to be broken at the canonical seesaw scale, $10^{14-15}$ GeV.  

Quark-lepton unification at scales not too far from to the electroweak symmetry breaking can be realized in the theoretical framework proposed
in Ref.~\cite{FileviezPerez:2013zmv}.
In this context, the neutrino masses are generated via the inverse seesaw mechanism and the gauge symmetry can be broken at the TeV scale. A similar model can be found in Ref.~\cite{Smirnov:1995jq}, but without the needed mechanism for neutrino masses that allow us to have a realistic theory at the low scale. The theory proposed in Ref.~\cite{FileviezPerez:2013zmv} is a realistic renormalizable theory that predicts the existence of leptoquark (LQ)
fields. These are fields that couple to simultaneously to a SM quark and lepton.
It predicts a vector LQ living in the off-diagonal part of the gauge field $SU(4)_C$ multiplet, and two scalar LQs with
couplings proportional to the quark and lepton masses.
This theory has been investigated in Refs.\cite{FileviezPerez:2022fni,FileviezPerez:2022rbk,FileviezPerez:2021arx,FileviezPerez:2021lkq,Flacke:2025xwl,FileviezPerez:2023rxn,Gedeonova:2022iac,Murgui:2021bdy,Miralles:2019uzg,Faber:2018qon}.
For a general review of LQ physics, see Ref.\cite{Dorsner:2016wpm}. 

In this article, we investigate in detail the predictions in the minimal quark-lepton unification theory~\cite{FileviezPerez:2013zmv}.
Our main interest is to understand the unique signatures at the Large Hadron Collider (LHC) and use the current available experimental
data to find the allowed parameter space.
We pay special attention at the properties of the right-handed neutrinos and LQ fields.
We find that some of the most important couplings of LQs to SM fermions are determined by the masses of the third generation fermions.
The right-handed neutrinos in this theory have interesting properties -- their decays via the LQs can dominate, giving rise to very
interesting signatures.
The LQs can also decay to right-handed neutrinos, changing the branching ratios to the SM fermions, with important
implications for experimental sensitivity.
We provide a detailed analysis of the production channels and all possible decays, and their potential impact on existing LHC measurements.
These results tell us that despite current measurements and searches, significant allowed parameter space for low
scale quark-lepton unification remains, and one could hope to test this theory in the near future.

This article is organized as follows: In Section~\ref{theory} we discuss the minimal theory for quark-lepton unification,
while in Section~\ref{LQs} we provide a detailed discussion of the different LQ fields.
In Section~\ref{collider} we present the predictions for the main collider signatures, discussing the production cross sections
and the different decay modes, and detemine constraints from current measurements.
Finally, we summarize our main findings in Section~\ref{summary}.
\section{LOW SCALE QUARK-LEPTON UNIFICATION}
\label{theory}
This minimal theory for matter unification is based on the gauge symmetry~\cite{FileviezPerez:2013zmv} $$\SU(4)_C \otimes \SU(2)_L \otimes \U(1)_R,$$ while the SM
matter fields are unified as follows
\begin{eqnarray}
F_{q_L} &=&
\left(
\begin{array}{cccc}
u_r & u_g & u_b  & \nu 
\\
d_r & d_g & d_b  & e
\end{array}
\right)_L \sim (\mathbf{4}, \mathbf{2}, 0), \\[1ex]
F_{u_R} &=&
\left(
\begin{array}{cccc}
u_r & u_g & u_b & \nu
\end{array}
\right)_R \sim (\mathbf{{4}}, \mathbf{1}, 1/2), 
\\[1ex]
 F_{d_R} &=&
\left(
\begin{array}{cccc}
d_r & d_g & d_b & e
\end{array}
\right)_R \sim (\mathbf{{4}}, \mathbf{1}, -1/2).
\end{eqnarray}
Here $u_i$ and $d_i$ (with $i=r,g,b$) are the quarks with different colors. 

The Lagrangian of the theory can be written as
\begin{eqnarray}
\mathcal{L} _{421} &=& - \frac{1}{2}  {\rm Tr} (V_{\mu \nu} V^{\mu \nu}) - \frac{1}{2}  {\rm Tr} (W_{\mu \nu} W^{\mu \nu})  - \frac{1}{4}  B_{R\mu \nu} B^{\mu \nu}_R  \nonumber \\
& + &  i \bar{F}_{q_L} \slashed{D} F_{q_L} +  i \bar{F}_{u_R} \slashed{D} F_{u_R} +  i \bar{F}_{d_R} \slashed{D} F_{d_R}  +  \mathcal{L}_Y - V(H,\chi,\Phi),
\end{eqnarray}
where 
\begin{equation}
V_{\mu \nu}= \partial_\mu V_\nu -  \partial_\nu V_\mu + i g_4 [V_\mu, V_\nu],
\end{equation}
is the strength tensor for the $\SU(4)_C$ gauge fields, $V_\mu \sim (\mathbf{15},\mathbf{1},0)$. The strength tensor for the $\SU(2)_L$ gauge fields, $W_\mu \sim (\mathbf{1},\mathbf{3},0)$, is given by
\begin{equation}
W_{\mu \nu}= \partial_\mu W_\nu -  \partial_\nu W_\mu + i g_2 [W_\mu, W_\nu],    
\end{equation} 
and for the $\U(1)_R$ gauge field, $B_{R\mu} \sim (\mathbf{1},\mathbf{1},0)$, we have $B_{R}^{\mu \nu}= \partial^\mu B_{R}^{\nu} -  \partial^{\nu} B_{R}^{\mu}$.  
The covariant derivatives for the 
fermionic fields are given by
\begin{eqnarray}
\slashed{D} F_{q_L} &=& \gamma^\mu (\partial_\mu + i g_4 V_\mu + i g_2 W_\mu) F_{q_L}, \\
\slashed{D} F_{u_R} &=&  \gamma^\mu (\partial_\mu + i g_4 V_\mu + \frac{i}{2}  g_R B_{R \mu} ) F_{u_R},   \\
\slashed{D} F_{d_R} &=&  \gamma^\mu (\partial_\mu + i g_4 V_\mu - \frac{i}{2}  g_R B_{R\mu} ) F_{d_R}.
\end{eqnarray}
The Yukawa interactions for the charged fermions can be written as
\begin{align}
	-\mathcal{L}_Y = Y_1 \bar{F}_{q_L} i \sigma_2 H_1^* F_{u_R} 
    + Y_2 \bar{F}_{u_R} \Phi^\alpha F_{q_L}^\beta \epsilon_{\alpha \beta} + Y_3 \bar{F}_{q_L} H_1 F_{d_R} + Y_4 \bar{F}_{q_L} \Phi F_{d_R} +  {\rm H.c.},
    \label{Yukawa}
\end{align}
where $H_1 \sim (\mathbf{1},\mathbf{2},1/2)$ and $\Phi \sim (\mathbf{15},\mathbf{2},1/2)$ are needed to generate fermion masses in a consistent manner. Here $\alpha$ and $\beta$ are $SU(2)_L$ indices.

Now, we can discuss in detail the properties of the different fields:
\begin{itemize}
\item {\textit{Gauge Fields}}: In this theory the $SU(4)_C$ gauge fields live in 
  \begin{eqnarray}
    V^\mu =
    \left(
    \begin{array} {cc}
      G^\mu & X^\mu/\sqrt{2}  \\
      (X^\mu)^*/\sqrt{2} & 0  \\
    \end{array}
    \right) + T_{15} \ V_{15}^{\mu} \sim (\mathbf{15}, \mathbf{1},0),
  \end{eqnarray}
  where $G^\mu \sim (\mathbf{8},\mathbf{1},0)_\text{SM}$ are the SM gluons, $X^\mu \sim (\mathbf{3},\mathbf{1},2/3)_\text{SM}$ are vector LQs, and $V_{15}^{\mu} \sim (\mathbf{1},\mathbf{1},0)_\text{SM}$. The $T_4$ generator of $\SU(4)_C$ in the above equation is normalized as
  \begin{equation}
    T_{15} =
    \frac{1}{2 \sqrt{6}} \rm{diag} (1,1,1,-3).    
  \end{equation}
\item {\textit{Higgs Sector}}: The Higgs sector is composed of three Higgses, $\Phi$ and $\chi$ and the SM Higgs $H_1$, with the following properties
  \begin{eqnarray}
    H_1 & = & 
    \left( \begin{array} {c}
      H^+_1  \\
      H^0_1 \\
    \end{array}
    \right)
    \sim (\mathbf{1}, \mathbf{2}, 1/2),\\
    \chi &=& \left(  \chi_r  \  \chi_g \ \chi_b \ \chi^0 \right) \sim (\mathbf{4}, \mathbf{1}, 1/2), 
  \end{eqnarray}
  and
  \begin{eqnarray}
    \Phi &=& 
    \left(
    \begin{array} {cc}
      \Phi_8 & \Phi_3  \\
      \Phi_4 & 0  \\
    \end{array}
    \right) + T_4 \ H_2 \sim (\mathbf{15}, \mathbf{2}, 1/2).
  \end{eqnarray}
  Here $H_2 \sim (\mathbf{1}, \mathbf{2}, 1/2)_\text{SM}$ is a second Higgs doublet, $\Phi_8 \sim (\mathbf{8}, \mathbf{2}, 1/2)_\text{SM}$,
  is a scalar octet and the scalar LQs are
  $\Phi_3 \sim (\mathbf{\bar{3}}, \mathbf{2},-1/6)_\text{SM}$ and  $\Phi_4 \sim (\mathbf{3}, \mathbf{2}, 7/6)_{\rm SM}$. 
  
  The gauge group is spontaneously broken to the SM gauge group by the vacuum expectation value (VEV) of the scalar
  field $\langle \chi^0 \rangle =v_{\chi}/\sqrt{2}$, which gives mass to the vector LQ $X_\mu$, defining the scale of matter unification.

  \item {\textit{Quark-Lepton Unification Scale:}}  
  Once $\chi$ gets a vev, the new massive vectors associated with the broken \mbox{generators} of $\SU(4)_C$ acquire mass. The mass of the vector LQs, $X_\mu \sim ({\bf{3}},1,2/3)_{\rm SM}$, is given by
  \begin{equation}
    M_{X}^2   = \frac{1}{4} g_4^2 v_\chi^2\,.
  \end{equation}
  The new neutral massive gauge boson, $Z'_\mu$, is a linear combination of the $V_{15\mu}$ gauge boson associated to the broken $\SU(4)$ generator $T_{15}$ and the $\text{U}(1)_R$ gauge boson, $B_{R\mu}$. Its mass comes from the $\chi$ kinetic term,
  \begin{equation}
    \begin{split}
      {\cal L}&\supset (D_\mu \chi)^\dagger (D^\mu \chi)  \\
      &\supset  \frac{v_\chi^2}{8} \begin{pmatrix}  V_{15\mu} & B_{R\mu}  \end{pmatrix} \begin{pmatrix}   \dfrac{3g_4^2}{2}  & - \dfrac{3g_R g_4 }{\sqrt{6}}  \\  -\dfrac{3 g_R g_4 }{\sqrt{6}} &g_R^2   \end{pmatrix} \begin{pmatrix}  V_{15}^\mu \\ B_R^\mu \end{pmatrix}\,.
    \end{split}
  \end{equation}
  The physical states are defined by
  \begin{equation}\label{eq:rotationS}
    \begin{pmatrix} V_{15\mu}  \\ B_{R\mu} \end{pmatrix} = \begin{pmatrix} \cos \theta_S & \sin \theta_S \\ -\sin \theta_S & \cos \theta_S \end{pmatrix} \begin{pmatrix}   Z_\mu' \\ B_\mu \end{pmatrix}\,,
  \end{equation}
  where the angle $\theta_S$ is given by the gauge couplings
  \begin{equation}
    \sin \theta_S =\frac{g_R}{\sqrt{g_R^2 + \tfrac{3}{2} g_4^2}}\,, \quad {\rm{and}} \quad \cos \theta_S = \frac{ g_4}{\sqrt{\tfrac{2}{3}g_R^2 + g_4^2}}\,.
  \end{equation}
  Notice that the hypercharge gauge coupling is given by $g_1 = g_R \cos \theta_S$, while the strong gauge coupling is $g_3 = g_4$. Using these relations one can write:
  \begin{equation}
    \sin^2 \theta_S = \frac{2}{3}\frac{\alpha_{\rm em}}{\alpha_s} \frac{1}{\cos^2\theta_W}\,.
  \end{equation}
  The mass of the gauge boson $Z'$ is given by 
  \begin{equation}
    M_{Z'}^2 = \frac{1}{4}\left (\frac{3}{2} g_4^2 + g_R^2\right )v_\chi^2\,.
  \end{equation}
  Therefore, one finds the following relation between the vector LQ and $Z^{'}$ masses:
  \begin{equation}\label{eq:MZMXratio}
    \frac{M_X^2}{M_{Z'}^2} = \frac{2}{3}\cos^2\theta_S.
  \end{equation}
  In this theory the idea of quark-lepton unification can be realized below the canonical seesaw scale, $10^{14-15}$ GeV. However, one faces strong bounds from lepton flavour violating processes.
  The vector-LQ mediates the rare process, $K_L^0 \to e^{\pm} \mu^{\mp}$, at tree-level and naively its mass should be
\begin{equation}
M_X \gtrsim 10^3 \ \rm{TeV}.    
\end{equation}
For details see Refs.~\cite{Valencia:1994cj,Gedeonova:2022iac,FileviezPerez:2021lkq,FileviezPerez:2022rbk}. It is important to mention that the unknown mixing between quarks and leptons are very important to predict the processes such as $K_L^0 \to e^{\pm} \mu^{\mp}$. Therefore, one could imagine scenarios where the quark-lepton unification can be below $10^3$ TeV.
 \item {\textit{Charged Fermion Masses}}: 
  The mass matrices for the SM charged fermions read as
  \begin{eqnarray}
    M_U &=& Y_1 \frac{ v_1}{\sqrt{2}} - \frac{1}{2 \sqrt{6}} Y_2^\dagger \frac{ v_2}{\sqrt{2}}, 
     \label{eq:fermionmasses1}\\
    M_D &=& Y_3  \frac{v_1}{\sqrt{2}} + \frac{1}{2 \sqrt{6}} Y_4 \frac{ v_2}{\sqrt{2}}, \\
    M_E &=& Y_3\frac{ v_1}{\sqrt{2}} - \frac{3}{2 \sqrt{6}} Y_4 \frac{ v_2}{\sqrt{2}}.
    \label{eq:fermionmasses2}
  \end{eqnarray}
  Here the VEVs of the Higgs doublets are defined as $\langle H^0_1 \rangle = v_1 / \sqrt{2}$, and $\langle H^0_2  \rangle  = v_2/\sqrt{2}$.
  Notice that without the scalar field $\Phi$, there would be only two independent Yukawa couplings, so the charged fermion
  masses could not be generated to be consistent with measurement. In our convention the charged fermion masses are diagonalized in the following way:
\begin{eqnarray}
U_L^\dagger M_U U_R &=& M_U^{diag}, \\
D_L^\dagger M_D D_R &=& M_D^{diag}, \\
E_L^\dagger M_E E_R &=& M_E^{diag}.
\end{eqnarray}
\item {\textit{Neutrino Masses}}: The Dirac mass term for neutrinos reads as:
\begin{equation}
M_\nu^D = Y_1 \frac{ v_1}{\sqrt{2}} + \frac{3}{2 \sqrt{6}} Y_2^\dagger \frac{ v_2}{\sqrt{2}}.
\label{eq:Dirac}
\end{equation}
Note that small neutrino masses would require extreme fine-tuning between the
  terms in Eq.~(\ref{eq:Dirac}).
  To generate small neutrino masses naturally at a low scale, one needs to go beyond the canonical seesaw mechanism.
  We can generate small Majorana  masses for the light neutrinos if we add three new singlet left-handed fermionic fields
  $S_L \sim (\mathbf{1}, \mathbf{1},0)$ and use the following interaction terms~\cite{FileviezPerez:2013zmv},
  \begin{eqnarray}
    - {\cal L}_{QL}^\nu &=&
    Y_5 \bar{F}_{u_R} \chi S_L  \ + \  \frac{1}{2} \mu S_L^T C S_L   + \mbox{h.c.}.
  \end{eqnarray} 
  In this case the mass matrix for neutrinos in the basis ($\nu$, $\nu^c$, $S$) reads as
  \begin{equation}
    \left( \nu \  \nu^c \  S  \right) 
    \left(\begin{array}{ccc} 
      0 & M_\nu^D & 0  \\ 
      (M_\nu^D)^T & 0 & M_\chi^D \\
      0 &  (M_\chi^D)^T & \mu
    \end{array}\right)  
    \left(\begin{array}{c} \nu \\  \nu^c \\ S  \end{array}\right).
  \end{equation}
  Here $M_\nu^D$ is given by Eq.~\eqref{eq:Dirac} and 
  $
  M_\chi^D = Y_5 \, v_\chi / \sqrt{2}.
  $
  Therefore, the light neutrino mass is given by 
  \begin{equation}
    m_\nu \approx \mu \, (M_\nu^D)^2 / (M_\chi^D)^2,
  \end{equation}
  when $M_\chi^D \gg M_\nu^D \gg \mu$ holds. Such a hierarchy is motivated by the different scales of the theory: $M_\chi^D \propto v_\chi$,
  which determines the scale of matter unification, $M_\nu^D \propto v_{1,2}$, which defines the electroweak scale,
  and $\mu$ is instead protected by a fermion symmetry, so that it is technically natural to assume it is small. When $M_\nu^D \sim 10 ^2$ GeV and $M_\chi^D \sim 10 ^6$ GeV , light neutrino mass constraint implies that the  $\mu $ has to be $\mathcal{O}(10^{-2})$ GeV.
  Notice that neutrinos would be massless in the limit $\mu \rightarrow 0$, which is the usual relation in the inverse seesaw mechanism. For the original proposal of the inverse seesaw mechanism see Ref.~\cite{Mohapatra:1986bd}.
  
\item {\textit{Higgs Decays}}: In Ref.~\cite{FileviezPerez:2022fni} the predictions for the Higgs decays in this theory were investigated.
  It was shown that the relation between the heavy CP-even $(H)$ and CP-odd Higgs $(A)$, decays can be used to test the idea of quark-lepton
  unification. In the case when there is non-flavor violation one finds the simple relations: 
  \begin{eqnarray}
    \Gamma (H \to \tau \bar{\tau}) &=& 3 \Gamma (H \to b \bar{b}), \\    \Gamma (A \to¯\tau \bar{\tau}) &=& 3 \Gamma (A \to b \bar{b}). 
  \end{eqnarray}
  for small values of $\tan \beta$, and 
  \begin{eqnarray}
    \Gamma (H \to \tau \bar{\tau}) &=& \Gamma (H \to¯b \bar{b})/3,\\
    \Gamma (A \to¯\tau \bar{\tau}) &=& \Gamma (A \to b \bar{b})/3. 
  \end{eqnarray}
  for large values of $\tan \beta$. These relations can be used to test the predictions for this simple theory for low scale quark-lepton
  unification. For more details see the discussion in Ref.~\cite{FileviezPerez:2022fni}. In the current study, we do not consider the
  phenomenological impact of the new Higgs doublet at collider energies.
\end{itemize}
\section{LEPTOQUARKS}
\label{LQs}
The theory predicts three types of LQ: one vector and two scalars.
Here we discuss the main interactions relevant for the study of the signatures at the LHC.
As mentioned above, the vector LQ mediates the rare process $K_L^0 \to e^{\pm} \mu^{\mp}$ at tree-level, and naively its mass should be above the $10^3$ TeV scale. Therefore, one cannot expect to observe the vector-LQ at the LHC unless this process is suppressed for other reasons. 
For this reason we focus on the signatures related to the scalar LQs predicted in this theory.
\begin{itemize}
\item $\Phi_3 \sim (\mathbf{\bar{3}}, \mathbf{2},-1/6)_\text{SM}$: The relevant Yukawa interactions for our study are given by
  \begin{equation}
    -{\cal L}  \supset  Y_2 \, \bar{\nu}_R \Phi_3^\alpha q_L^\beta \epsilon_{\alpha \beta} \ + Y_4 \, \bar{\ell}_L  \Phi_3 d_R +  {\rm H.c.}\, . 
  \end{equation}
  The scalar LQ $\Phi_3$ can be written in $\SU(2)_L$ components as,
  \beq
  \Phi_3 =  \mqty(\phi_3^{1/3} \\[1ex]\phi_3^{-2/3} ),
  \eeq
  where the numbers denote the electric charge. Notice that the two component of the $\Phi_3$ LQ have similar masses. Their 
  mass splitting is below a few hundred GeVs. For simplicity, 
  we will assume that they have the same mass.   
  
  The above interactions can be written explicitly as
  \begin{equation}
    -{\cal L}  \supset  Y_2 \, \bar{\nu}_R \phi_3^{1/3} d_L - Y_2 \bar{\nu}_R \phi_3^{-2/3} u_L \ + Y_4 \, \bar{\nu}_L  \phi_3^{1/3} d_R + Y_4 \bar{e}_L \phi_3^{-2/3} d_R \ + \ {\rm H.c.}\, . 
  \end{equation}
\item $\Phi_4 \sim (\mathbf{3}, \mathbf{2},7/6)_\text{SM}$: This scalar LQ has the following interactions with the SM fermions:
  \begin{equation}
    -{\cal L}  \supset Y_2 \, \bar{u}_R  \Phi_4^\alpha \ell_L^\beta \epsilon_{\alpha \beta} + \  Y_4 \, \bar{q}_L  \Phi_4 e_R +  {\rm H.c.}\, . 
  \end{equation}
  The $\Phi_4$ field can be written in $\SU(2)_L$ components as,
  \beq
  \Phi_4 =\mqty(\phi_4^{5/3} \\[1ex]\phi_4^{2/3} ).
  \eeq 
  Notice that both components will have similar masses.
  Using this equation we can write the $\Phi_4$ Yukawa interactions explicitly as
  \begin{equation}
    -{\cal L}  \supset Y_2 \bar{u}_R \phi_4^{5/3} e_L - Y_2  \bar{u}_R \phi_4^{2/3} \nu_L  +  
    Y_4 \bar{u}_L \phi_4^{5/3} e_R + Y_4 \bar{d}_L \phi_4^{2/3} e_R + {\rm H.c.}.
  \end{equation}
\end{itemize}
In this theory, neutrino masses can be small even when $Y_2 \to 0$, due to the inverse seesaw mechanism, but the entries in $Y_4$ cannot be arbitrarily
small because one needs a realistic relation between the masses of the down-like quarks and the charged leptons. The Yukawa matrix $Y_4$ can be written as:
\begin{equation}
Y_4=\sqrt{3} \frac{(M_D - M_E)}{v \sin \beta}=\sqrt{3} \frac{(D_L M_D^{diag} D_R^\dagger - E_L M_E^{diag} E_R^\dagger )}{v \sin \beta}.
\label{eq:Y4}
\end{equation}
Here $v=\sqrt{v_1^2+v_2^2}=246$ GeV and $\tan \beta=v_2/v_1$. 
The Yukawa matrix $Y_4$ defines the difference between the mass matrices for down quarks and charged leptons.
Therefore, the scalar LQ masses proportional to $Y_4$ will be related to the down and charged lepton masses. 

Using the above equation, we can write the Yukawa couplings proportional to $Y_4$ in as follows:
\begin{eqnarray}
\bar{d}_L Y_4 e_R \phi_4^{2/3}&\to& \frac{\sqrt{3}}{v \sin \beta} \ \bar{d} \left( M_D^{diag} V_R - V_L M_E^{diag}\right) P_R \ e \ \phi_4^{2/3},  
\end{eqnarray}
where the mixing matrices, $V_R=D_R^\dagger E_R$ and $V_L=D_L^\dagger E_L$, and $P_R=(1+\gamma_5)/2$.
The other interactions read as
\begin{eqnarray}
  \bar{e}_L Y_4 d_R \phi_3^{-2/3} & \to & \frac{\sqrt{3}}{v \sin \beta} \bar{e} \left( V_L^\dagger M_D^{diag} - M_E^{diag} V_R^\dagger \right) P_R \ d \ \phi_3^{-2/3},    \\
  \bar{u}_L Y_4 e_R \phi_4^{5/3} &\to& \frac{\sqrt{3}}{v \sin \beta} \bar{u} \left( V_{UD} M_D^{diag} V_R - V_{UD} V_L M_E^{diag} \right) P_R \ e \ \phi_4^{5/3},
\end{eqnarray}
with $V_{UD}=K_1 V_{CKM} K_2$ and $V_{EN}=K_3 V_{PMNS}$, where $K_i$ are diagonal unitary complex matrices. Finally, one has
\begin{eqnarray}
  \bar{\nu}_L Y_4 d_R \phi_3^{1/3} &\to& \frac{\sqrt{3}}{v \sin \beta} \bar{\nu} \ V_{EN} \left( V_L^\dagger M_D^{diag}-M_E^{diag} V_R^\dagger \right) P_R \ d \ \phi_3^{1/3}.
\label{eq:y4diag}
\end{eqnarray}
Notice that when we neglect the unknown mixing angles, $V_i \sim 1$, except for $V_{EN}$, the above equations tell us that the
scalar LQs couple mainly to the third generation of quarks and leptons.
Therefore, the main decays determined by the coupling $y_4$
\begin{equation}
y_4=\frac{\sqrt{3}}{v \sin \beta} (m_b - m_\tau),
\label{eq:y4}
\end{equation}
corresponding to $y_{4~33}^{LR}$ in the notation of \cite{Dorsner:2016wpm}, are
\begin{equation}
    \phi^{2/3}_4 \to \bar{\tau} b, \ \phi_3^{-2/3} \to \bar{b} \tau, \ \phi_4^{5/3} \to \bar{\tau} t, \ \phi_3^{1/3} \to \bar{b} \nu_i.
\end{equation}
{\textit{This very simple prediction tells us that some of the scalar LQs signatures are determined by the couplings to the third generation of quarks and leptons.}}

Using the above interactions one can find the following approximate relations between the decay widths:
\begin{eqnarray}
\frac{\Gamma (\phi_4^{2/3} \to \bar{\tau} b)}{\Gamma (\phi_4^{5/3} \to \bar{\tau} t)} &\simeq& \frac{M_{\phi_4^{2/3}}}{M_{\phi_3^{5/3}}}, \ \frac{\Gamma (\phi_4^{2/3} \to \bar{\tau} b)}{\Gamma (\phi_3^{-2/3} \to \bar{b} \tau)} \simeq \frac{M_{\phi_4^{2/3}}}{M_{\phi_3^{-2/3}}} ,\
\frac{\Gamma (\phi_4^{2/3} \to \bar{\tau} b)}{\Gamma (\phi_3^{1/3} \to \bar{b} \nu)} \simeq \frac{M_{\phi_4^{2/3}}}{M_{\phi_3^{1/3}}},
\end{eqnarray}
we can use to test the predictions of this theory.
Unfortunately, this theory does not predict the $Y_2$ Yukawa coupling. 
In fact, $Y_2$ can be zero and still we can have a realistic theory for fermion masses. Assuming the largest coupling in $Y_2$ is to the third generation, one can predict some interesting relations, neglecting the fermion masses:
\begin{eqnarray}
\frac{\Gamma (\phi_3^{1/3} \to N_3 \bar{b})}{\Gamma (\phi_3^{-2/3} \to N_3 \bar{t})}\simeq\frac{M_{\phi_3^{1/3}}}{M_{\phi_3^{-2/3}}}, \
\frac{\Gamma (\phi_4^{5/3} \to t \bar{\tau})}{\Gamma (\phi_4^{2/3} \to t \bar{\nu}_\tau)}\simeq\frac{M_{\phi_4^{5/3}}}{M_{\phi_4^{2/3}}}, \
\frac{\Gamma (\phi_3^{-2/3} \to N_3 \bar{t})}{\Gamma (\phi_4^{2/3} \to t \bar{\nu}_\tau)}\simeq
\frac{M_{\phi_3^{-2/3}}}{M_{\phi_4^{2/3}}}.
\end{eqnarray}
In the scenarios when $Y_2$ is non-zero, for simplicity, we assume $Y_2 = \text{diag}(0,0,y_2)$, where $y_2$
corresponding to $y_{2~33}^{LR}$ in the notation of Ref.~\cite{Dorsner:2016wpm}.
It is important to emphasize that this simple theory for low scale quark-lepton unification tells us that the scalar LQs could have masses at the TeV scale and their decays proportional to $Y_4$ are determined by the couplings to the third generation of quarks and leptons.
\section{LHC PHENOMENOLOGY}
\label{collider}
\subsection{Production Channels}
\begin{figure}[t]
    \centering
    \begin{subfigure}[b]{0.48\textwidth}
        \centering
        \includegraphics[width=\textwidth]{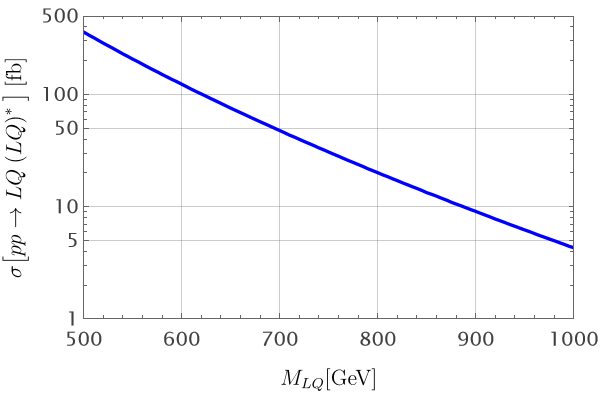}
        \caption{}
        \label{subfig:LQ_pair}
    \end{subfigure}  
    \begin{subfigure}[b]{0.47\textwidth}
        \centering
        \includegraphics[width=\textwidth]{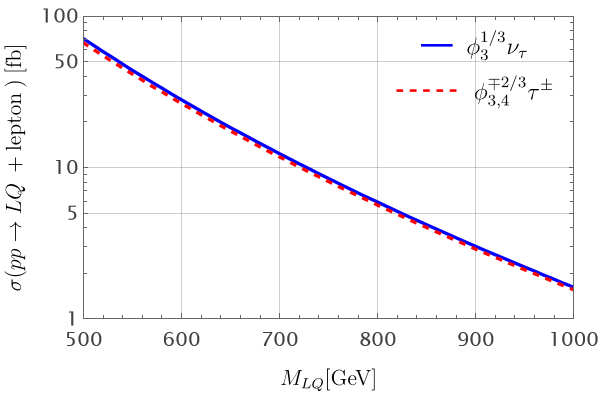}
        \caption{}
        \label{subfig:LQ_lepton}
    \end{subfigure}          
    \par\vspace{0.2cm} 
    \caption{ a)  QCD Pair production cross section for the scalar LQ 
      at $\sqrt{s}= 13$ TeV, as a function of LQ mass.
      b) Production cross section of a scalar LQ and SM lepton as a function of LQ mass when $y_4=1$.
      The cross-section was computed using our UFO model and \madgraph ~\cite{Alwall_2014}.  }
    \label{fig:LQcross}
 \end{figure}

Over much of the parameter range of the theory, the dominant production process is pair production via the strong coupling
of the LQs, 
\begin{eqnarray}
pp &\to & LQ \ {LQ}^{*}, \\
pp &\to & \textrm{lepton} \ + \  LQ,  
\end{eqnarray}
where the incoming partons can be $gg$ or $q\bar{q}$.
The pair production cross section for smaller LQ masses and large values of the Yukawa coupling, single production in association with a SM lepton can also be significant. To do calculate the cross sections and decay modes for the theory, we have implemented the model in \texttt{FeynRules}~\cite{Alloul:2013bka}
and exported it as a UFO~\cite{Degrande:2011ua} file which is then read by Monte Carlo event generators,
in particular here \madgraph\cite{Alwall_2014}
and \herwig~\cite{Bellm:2015jjp}, to simulate LQ production and decay at the LHC in proton-proton collisions at 13~TeV.
Examples of the cross-section dependence on LQ mass are shown in Fig.~\ref{fig:LQcross}. Fig.~\ref{subfig:LQ_pair} shows the leading order (LO) QCD pair production of the scalar LQ at $\sqrt{s}=13$ TeV, as a function of LQ mass. As all the scalar LQs couple with gluons in the same way, the QCD pair production cross section is the same for all of them.   In  Fig.~\ref{subfig:LQ_lepton}, we show the production cross section of a scalar LQ in association with a SM lepton, evaluated at $\sqrt{s}=13$ TeV for $y_4=1$.  The solid blue and the red dashed lines show the production cross section for $\phi_3^{1/3} \nu_\tau$ and $\phi_{3,4}^{\mp 2/3} \tau^{\pm}$ respectively. The single production cross section of a LQ depends on the Yukawa coupling and can be significant for a large value of the Yukawa coupling.
\subsection{Decay Modes and Signatures}
The decay modes of all the LQs will be dominantly to third-generation SM fermions or, if allowed,
to the heavy right-handed neutrino, which may then itself decay via diagrams involving an off-shell LQ,
discussed in Section~\ref{sec:Ndecay}. The decays to first- and second-generation fermions are
also implemented in our Feynrules model, but are suppressed according to Eqs.~(\ref{eq:y4diag}) and~(\ref{eq:y4}).

\begin{figure}[t]
  \centering
  \begin{subfigure}[b]{0.47\textwidth}
    \centering
    \includegraphics[width=\textwidth]{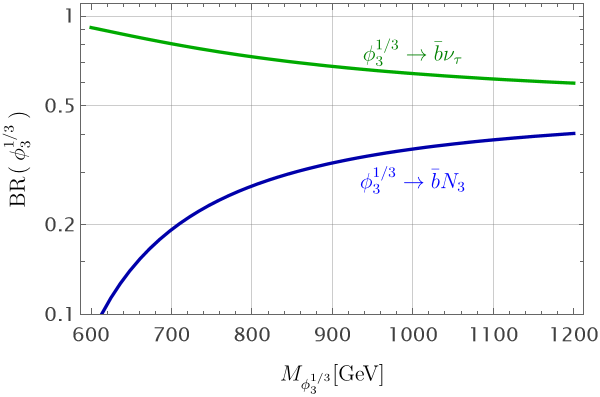}
    \caption{}
    \label{subfig:BRphi13}
  \end{subfigure}  
  \begin{subfigure}[b]{0.47\textwidth}
    \centering
    \includegraphics[width=\textwidth]{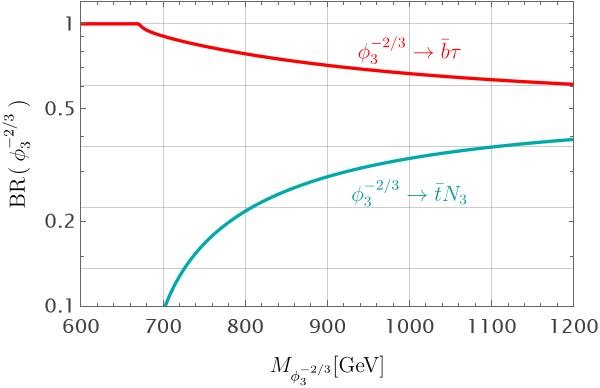}
    \caption{}
    \label{subfig:BRphi23N}
  \end{subfigure}          
  \begin{subfigure}[b]{0.47\textwidth}
    \centering
    \includegraphics[width=\textwidth]{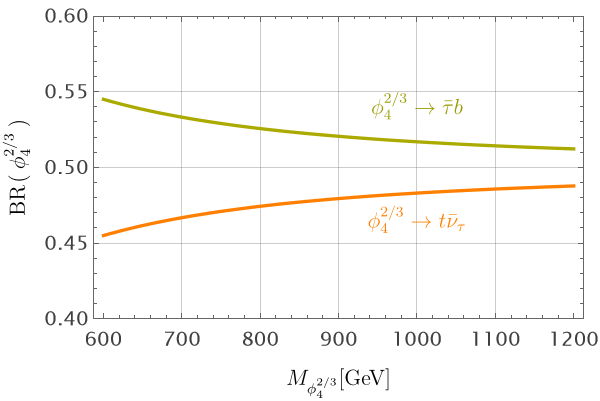}
    \caption{}
    \label{subfig:BRphi23P}
  \end{subfigure}          
  \par\vspace{0.2cm}
  \caption{Branching ratios of scalar LQ decays as a function of LQ mass. Fig. 2a) shows the Branching ratios of $\phi_3^{1/3}$ and
    Fig. 2b) shows the Branching ratios of $\phi_3^{-2/3}$. Fig. 2c) shows the branching ratios of $\phi_4^{2/3}$.
    Here, for illustration, we assumed $M_N = 500$ GeV, $y_4=1$, and $y_2=1$ in all scenerios.  }
  \label{fig:detector_signatures}
\end{figure}

The dominant decays of each LQ are as follows.
\begin{itemize}
\item{$\phi_3^{1/3}$:}
  When $y_2=0$, the scalar LQ $\phi_3^{1/3}$ decays mainly into a $b$-quark and neutrino, while when $y_2 \neq 0$, it may additionally decay into a
  right-handed neutrino $N$ and a $b$-quark. Therefore, the main signatures are:
  \begin{equation}
    pp \to \phi^{1/3}_3 (\phi^{1/3}_3)^{*} \to  \bar{b}\nu (b\bar{\nu}),  \bar{b}N (b\bar{N}).
    \label{eq:phi13decay}
  \end{equation}
  The branching ratio dependence on mass is shown in Fig.~\ref{subfig:BRphi13} for $\phi^{1/3}_3$ for example values
  $M_N = 500$ GeV, $Y_4=\text{diag}(0,0,1)$, and $Y_2=\text{diag}(0,0,1)$.
  \end{itemize}
  \begin{itemize}
\item{$\phi_3^{-2/3}$:}
  When $y_2=0$, the scalar LQ $\phi_3^{-2/3}$ decays mainly into a $b$-quark and $\tau$-lepton, while when $y_2 \neq 0$, it may additionally decay into a
  right-handed neutrino $N$ and a top-quark. Therefore, the main signatures are:
  \begin{equation}
    pp \to \phi^{-2/3}_3 (\phi^{-2/3}_3)^{*} \to  \bar{b}\tau (b\bar{\tau}) , \bar{t}N (t\bar{N}).  
    \label{eq:phi23Ndecay}
  \end{equation}
  The branching ratio dependence on mass is shown in Fig.~\ref{subfig:BRphi23N} for $\phi^{-2/3}_3$ for
  the same example values of $M_N, y_4$ and $y_2$.
\item $\phi_4^{2/3}$:
  When $y_2=0$, the scalar LQ $\phi_4^{2/3}$ decays mainly into a $b$-quark and $\tau$-lepton, while when $y_2 \neq 0$, it may additionally decay into a
  SM neutrino $\nu$ and a top-quark. Therefore, the main signatures are:
  \begin{equation}
    pp \to \phi^{2/3}_4 (\phi^{2/3}_4)^{*} \to b \bar{\tau} (\bar{b}\tau),  t\bar{\nu} (\bar{t}\nu).  
    \label{eq:phi23Pdecay}
  \end{equation}
  The branching ratio dependence on mass is shown in Fig.~\ref{subfig:BRphi23P} for $\phi^{2/3}_4$ for
  the same example values of $M_N, y_4$ and $y_2$.
\item $\phi_4^{5/3}$:
  The dominant decay channel of the $\phi^{5/3}_4$ scalar leptopquark, regardless of the value of $y_2$, is
  \begin{equation}
    pp \to \phi^{5/3}_4 (\phi^{5/3}_4)^{*} \to t \bar{\tau} (\bar{t} \tau).
    \label{eq:phi53Ndecay}
  \end{equation}

\end{itemize}
\subsubsection{Heavy Neutrino Decays}
\label{sec:Ndecay}
\begin{figure}[t]
    \centering
    \begin{subfigure}[b]{0.47\textwidth}
        \centering
        \includegraphics[width=\textwidth]{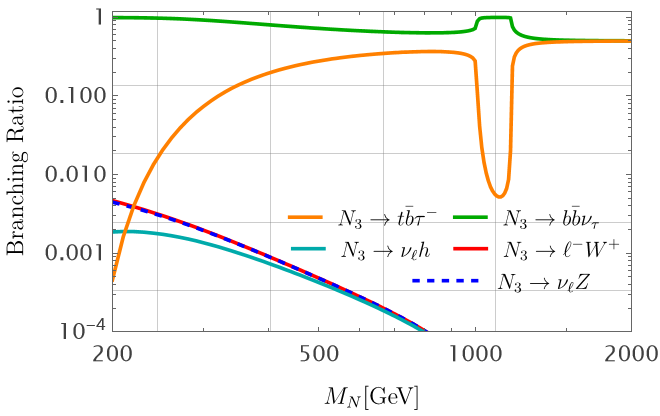}
        \caption{}
        \label{subfig:RN1}
    \end{subfigure}  
    \begin{subfigure}[b]{0.47\textwidth}
        \centering
        \includegraphics[width=\textwidth]{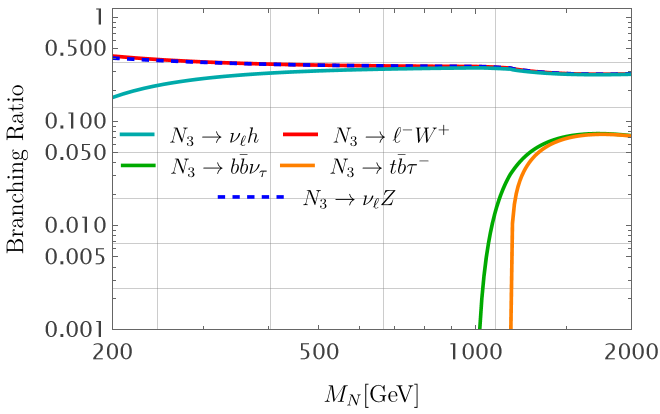}
        \caption{}
        \label{subfig:RN2}
    \end{subfigure}          
    \par\vspace{0.1cm} 
   \caption{Branching ratios of 3rd generation right-handed neutrinos as a function of mass.  Here, 
      we assumed  a) $y_2 = y_4 =1$ and 
      b) $y_2 = y_4 = 0.01$. In both scenarios, we fix $M_{\phi^{1/3}_3}=M_{\phi^{-2/3}_3}=1$ TeV and assume the mixing between the SM neutrinos and right-handed neutrinos to be $V_{\nu N}=10^{-4}$.}
    \label{fig:RNdecay}
 \end{figure}

The right-handed neutrinos can decay via the mixing with the SM neutrinos and can have 3-body decays mediated by the scalar LQs 
$\phi_3^{1/3}$ and $\phi_3^{-2/3}$. Within the framework of the inverse seesaw mechanism, the mixing between SM neutrinos and the right handed neutrinos can be written as 
\begin{eqnarray}
    V_{\nu N} \simeq \sqrt{\frac{m_\nu}{\mu}}.
\end{eqnarray}
When $\mu \sim 10 ^{-2}$ GeV, the mixing between the right-handed neutrinos and SM neutrinos is $\mathcal{O}(10^{-4})$. In Fig.~\ref{subfig:RN1} we show the branching ratios of 3rd generation right-handed neutrinos as
a function of its mass assuming $y_2 = y_4=1$ and $M_{\phi^{1/3}_3}=M_{\phi^{-2/3}_3}=1$ TeV.
 In this scenario, right-handed neutrino decays into $\nu_\ell h , \hspace{0.1 em} \nu_\ell Z, \hspace{0.1 em} \ell^{\pm}W^{\mp}$  via the mixing with the SM neutrinos, and these processes are suppressed due to the small mixing angle. The striking behaviour in the decays mediated via the LQs can be understood in terms of the available phase-space for the decay.
At $M_{N}<1$~TeV both LQs are virtual, and the top mass suppresses the $t\bar{b}\tau$ decay.
The significance of this decay rises with $M_{N}$, but is then further suppressed relative to the $b\bar{b}\nu_{\tau}$ decay
around 1~TeV, where the $\phi^{1/3}_3$ LQ can be produced on-shell,
$N_3 \rightarrow \phi^{1/3}_3 b \rightarrow \nu_{\tau}b\bar{b}$, while there is not enough
energy to produce $\phi^{-2/3}_3 t$ on-shell.
At higher $M_{N}$, phase-space effects become irrelevant and both decays are equally likely. In Fig.~\ref{subfig:RN2}, we show the branching ratios for 3rd generation right-handed neutrinos when $y_2 =y_4 = 0.01$. In this case, when the neutrino mass is below 1000 GeV, both LQs remain off-shell, and due to small Yukawa couplings, the decays mediated via the LQs are suppressed compared to the decays arising from the mixing with SM neutrinos. At larger masses, these decay branching ratios become comparable.
Over the parameters of the theory considered here, the decays via LQs are the most phenomenologically relevant,
since they generally dominate when $Y_2$ is significant, and when $Y_2$ is small, the production of heavy neutrinos is
itself suppressed.


\subsection{Sensitivity of LHC measurements}

Many of the differential cross sections which have already been measured by the LHC experiments have, potentially at least,
sensitivity to signatures of the type discussed above.
The \contur~\cite{Butterworth:2016sqg,Buckley:2021neu} application is designed to
allow a rapid check of detailed models against those measurements which are available in Rivet~\cite{Buckley:2010ar}.
Events are simulated using Herwig, based upon the UFO files of the theory.
The simulated events are then passed through \rivet~4~\cite{Bierlich:2024vqo} and analysed with
\contur~3.1~\cite{CONTUR:2025yis}.
This involves injecting the LQ contribution on top of the SM predictions for a wide range
of LHC measurements, and evaluating the ratio between the likelihood evaluated for the SM alone given the data,
and the likelihood for SM+LQ contributions, .
The $CL_S$ method~\cite{Read:2002hq} is used to determine an exclusion limit;
the expected exclusion limit is also determined by setting the data to the SM values but preserving its uncertainties.

We study each of the two LQ doublets $\Phi_3$ and $\Phi_4$ in turn, assuming the LQs within each doublet
are degenerate in mass. In each case the other doublet is decoupled by setting the mass 10~TeV. After that, we consider the
case where both doublet masses can be around the TeV scale.

\subsubsection{Limits and sensitivity to $\Phi_3$}
We first study the $\Phi_3$ doublet, which contains the $\phi^{-1/3}_3$ and $\phi^{-2/3}_3$ LQs, assuming the $\Phi_4$ is decoupled.
We begin by considering the case $y_2=0$.
In this case, the relevant free parameters are $y_4$ which we allow to vary in the range $0.01 \leq y_4 \leq 1$, and the LQ mass $M_{\Phi_3}$ which
we scan in the range 500~GeV to 1.5~TeV. The heavy neutrino mass is set to $M_N=10$~GeV,  although with $y_2=0$ this has no influence,
since the decay is always to SM particles -- $b$-quarks, and SM neutrinos or $\tau$-leptons, see Eqs.~(\ref{eq:phi13decay}) and~(\ref{eq:phi23Ndecay}).
\begin{figure}[t]
  \centering
      \includegraphics[width=0.80\linewidth]{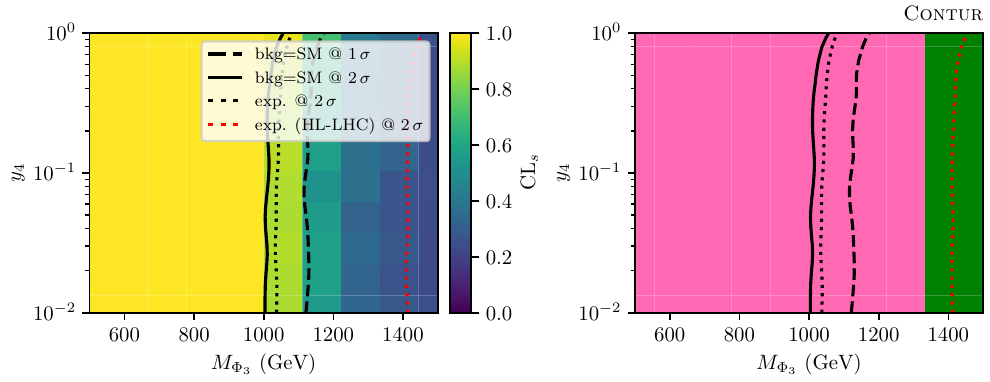}
      \caption{\contur exclusion plot of $M_{\Phi_3}$ against $y_4$ at $y_2=0$. The solid black line indicates the 95\% exclusion
        and the dashed black line the 68\% exclusion. The dotted black line shows the expected exclusion, where the data
        exactly coincide with the SM, and the dotted red line is an estimate of the eventual sensitivity after 3~ab$^{-1}$ of
        integrated luminosity.}
  \begin{tabular}{llll}
    \swatch{hotpink}~$\tau^+\tau^-$ \cite{ATLAS:2025oiy} & 
    \swatch{green}~\MET{}+jet \cite{ATLAS:2024vqf} & 
  \end{tabular}
  \label{fig:m3_y4_y2is0}
\end{figure}
The result is shown in Fig.~\ref{fig:m3_y4_y2is0}.
The left-hand plot indicates that masses below about 1~TeV are excluded at the 95\% c.l., (solid black line)
and the right-hand plot shows that this is dominantly
due to di-$\tau$ and missing transverse momentum signatures, which are incompatible
with the relevant ATLAS measurements~\cite{ATLAS:2024vqf,ATLAS:2025oiy}. In each of the
exclusion plots discussed, we also show the 68\%  (i.e. $1\sigma$) exclusion and the expected exclusion (if the data were to precisely agree with the SM).
Also shown is a crude estimate of the potential reach of the high-luminosity LHC, obtained by scaling the experimental uncertainties
of the current measurements according to a target luminosity of 3000~fb$^{-1}$. This is a conservative estimated since it does not take into
account any extension of the likely of measurements into new phase space.

There is also no strong dependence on $y_4$, since the
production cross section is dominated by pair production via the strong interaction. A slight strengthening of the constraints
as $y_4$ approaches unity is due to the influence of single LQ production.

\begin{figure}[t]
  \centering
    \begin{subfigure}[t]{0.48\linewidth}
      \includegraphics[width=\linewidth]{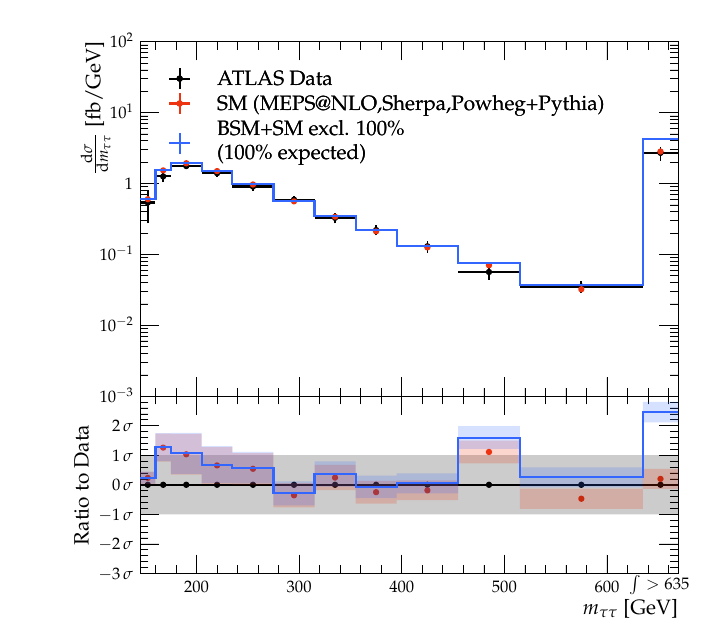}
      \caption{\label{fig:ditaurivet}}
    \end{subfigure}
    \begin{subfigure}[t]{0.48\linewidth}
      \includegraphics[width=\linewidth]{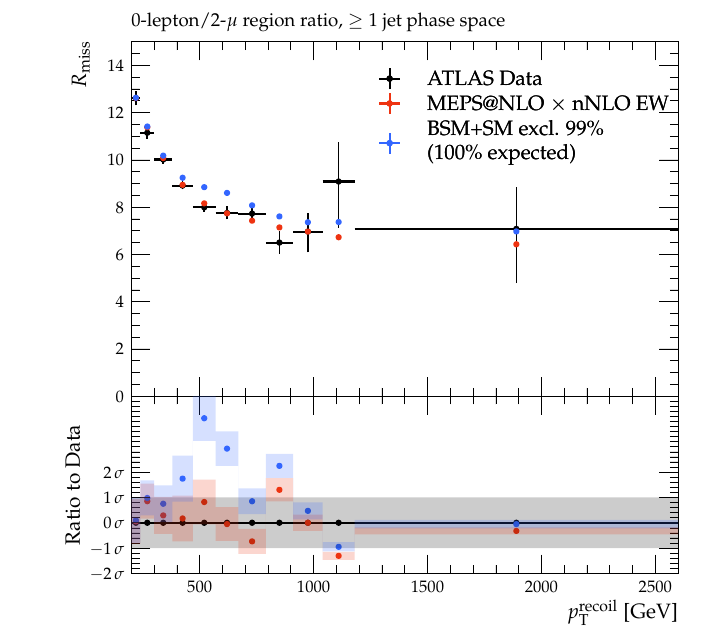}
      \caption{ \label{fig:metrivet}}
    \end{subfigure}
    \caption{ a) The di-$\tau$ invariant mass measurement (see text) showing the predicted signal for the point in Fig.~\ref{fig:m3_y4_y2is0} with
        $y_4 = 0.13$ and $M_{\Phi_3} = 944$~GeV. 
         b) The recoil \pT ratio (see text) showing the predicted signal for the point in Fig.~\ref{fig:m3_y4_y2is0} with
        $y_4 = 0.13$ and $M_{\Phi_3} = 600$~GeV.}
\end{figure}

An example plot showing the ATLAS measurement of the differential cross section with respect to the di-$\tau$ invariant mass~\cite{ATLAS:2025oiy} 
is shown in Fig.~\ref{fig:ditaurivet}.
The measurement is compared to the SM prediction from Sherpa~\cite{Sherpa:2019gpd},
with the top contribution from POWHEG+PYTHIA~\cite{Alioli:2010xd,Sjostrand:2007gs}, calculated at Next-to-leading order in QCD and
matched to leading-logarithmic parton showers (MEPS@NLO).
It can be seen that the signal would have appeared in the highest mass (overflow) bin.
Fig.~\ref{fig:metrivet} shows similarly an example of sensitivity in the ATLAS measurement
of the ratio ($R_{\textrm{miss}}$) of the differential cross sections with respect to the hadronic recoil \pT in zero lepton
events to two-muon events~\cite{ATLAS:2024vqf}, for a lower LQ mass.
The measurement is compared to the SM prediction from ~\cite{Sherpa:2019gpd}, including higher-order electroweak corrections.
The signal has an impact principally between 500 and 800~GeV.

\begin{figure}[t]
  \centering
    \begin{subfigure}[t]{\linewidth}  
      \includegraphics[width=0.80\linewidth]{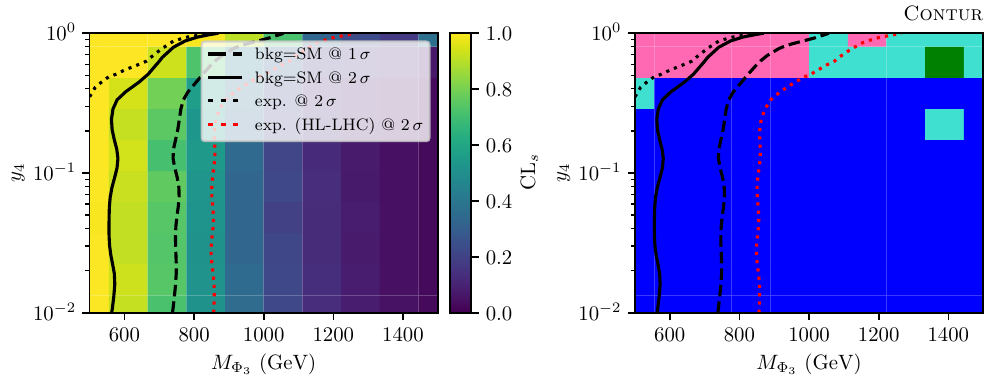}
      \caption{\label{fig:m3_y4_y2is1_mNis10}}
    \end{subfigure}
    \begin{subfigure}[t]{\linewidth}  
      \includegraphics[width=0.80\linewidth]{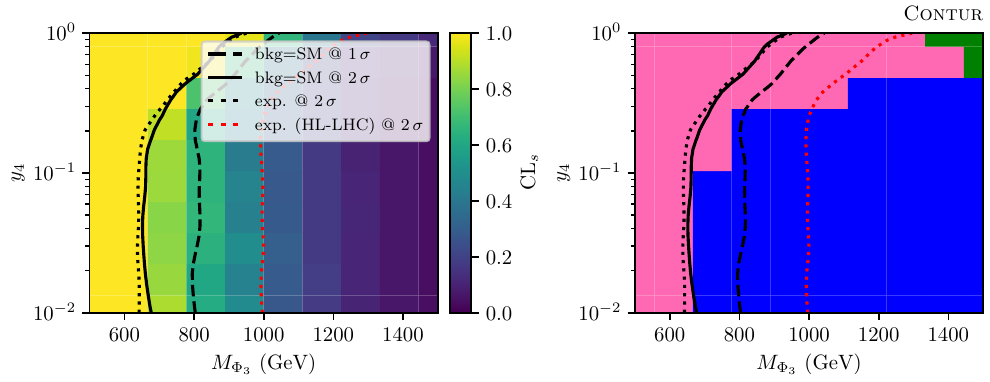}
      \caption{\label{fig:m3_y4_y2is1_mNis500}}
    \end{subfigure}
    \begin{subfigure}[t]{\linewidth}  
        \includegraphics[width=0.80\linewidth]{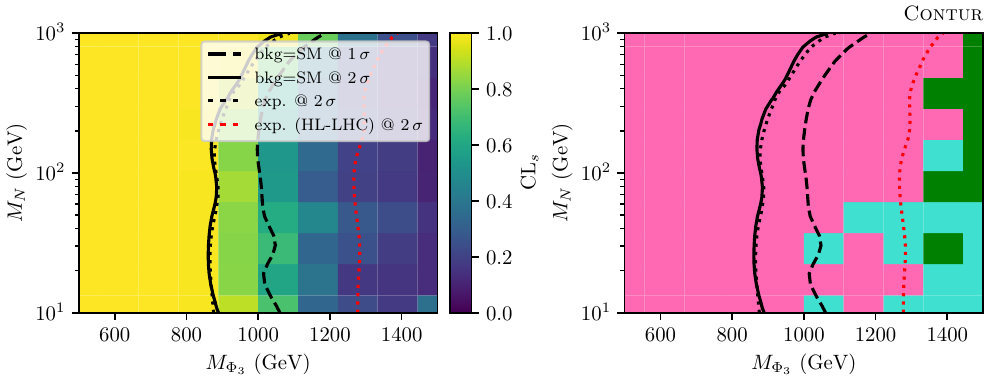}
        \caption{\label{fig:m3_mN_y2y4is1}}
    \end{subfigure}
    \caption{\contur exclusion plot of $M_{\Phi_3}$ against $y_4$ at $y_2=1$ and (a) $m_N = 10$ GeV (b) $m_N = 500$ GeV.
      (c) shows the dependence on $M_N$ for the case $y_4 = y_2 = 1$.  Lines as Fig.\ref{fig:m3_y4_y2is0}.
    \label{fig:phi13y2is1}}
    \begin{tabular}{llll}
        \swatch{turquoise}~$\ell_1\ell_2$+\MET{}+jet \cite{ATLAS:2024aht} & 
        \swatch{hotpink}~$\tau^+\tau^-$ \cite{ATLAS:2025oiy} & 
        \swatch{blue}~$\ell$+\MET{}+jet \cite{CMS:2021vhb,ATLAS:2022xfj} & 
        \swatch{green}~\MET{}+jet \cite{ATLAS:2024vqf} \\      
    \end{tabular}
    \label{m3_y4_y2is1}
\end{figure}

Allowing $y_2 \neq 0$ has a significant impact. The decays to the heavy neutrino, $N$, are now open, and this introduces a dependence on $M_N$ and on the decays of $N$. This is shown in Fig.~\ref{fig:m3_y4_y2is1_mNis10}. Here $M_N=10$~GeV, and for $y_4 < 1$, the $\phi^{1/3}_3$ decays mainly
to the heavy neutrino, $N$, and a $b$-quark, Eq.~(\ref{eq:phi13decay}).
Similarly, $\phi^{-2/3}_3$ can now decay to top quarks and $N$, Eq.~(\ref{eq:phi23Ndecay}).
The $N$ subsequently decays via an off-shell LQ to $b$-quarks and SM neutrinos or to top quarks, $b$-quarks and $\tau$-leptons,
as illustrated in Fig.~\ref{fig:RNdecay} for an example value of $M_{\Phi_3}$.
This greatly dilutes the \MET signature, leaving only a small residual sensitivity, and also 
reduces the sensitivity in $\tau$ channels, since the $\tau$s produced in $N$ decays generally
have lower \pT than those produced as primary products of LQ decay.
Some sensitivity is now present in top measurements, but this does not compensate for the reduction in other
channels and the overall sensitivity is reduced to around 550~GeV for most of the $y_4$ range.

The LQ decays to $N$ are somewhat suppressed when $M_N$ is larger, and
the sensitivity improves a bit, as shown in Fig.~\ref{fig:m3_y4_y2is1_mNis500}.
Fig.~\ref{fig:m3_mN_y2y4is1} shows the dependence on $M_N$ for $y_2 = y_4 = 1$. Only when $M_N$ approaches $M_{\Phi_3}$
is the senstivity seen for $y_2 = 0$ fully recovered, since in this case the decays to $N$ are kinematically
suppressed.
\subsubsection{Limits and sensitivity to $\Phi_4$}
Similar scans are performed for $M_{\Phi_4}$, under the assumption that the $\Phi_3$ is decoupled.
As can be seen from Eqs.~(\ref{eq:phi23Pdecay}) and~(\ref{eq:phi53Ndecay}), in this case the heavy
neutrino plays no role.

\begin{figure}[t]  
  \centering
  \begin{subfigure}[t]{\linewidth}  
    \includegraphics[width=0.80\linewidth]{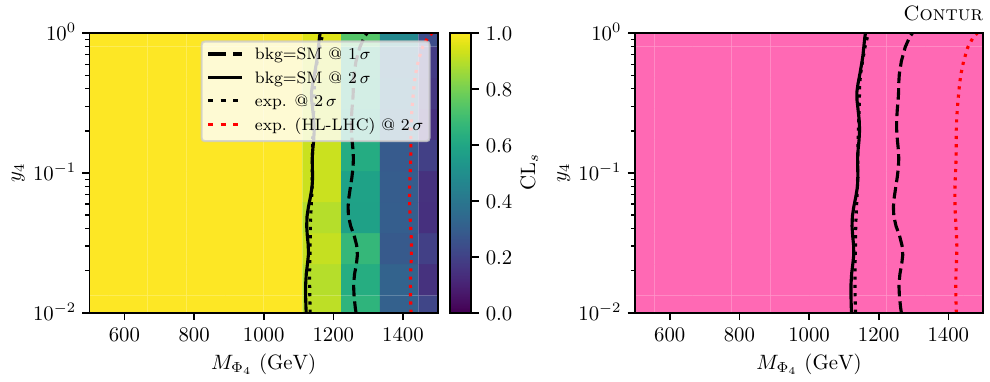}
    \caption{\label{fig:m4_y4_y2is0_mNis500}}
  \end{subfigure}
  \begin{subfigure}[t]{\linewidth}  
    \includegraphics[width=0.80\linewidth]{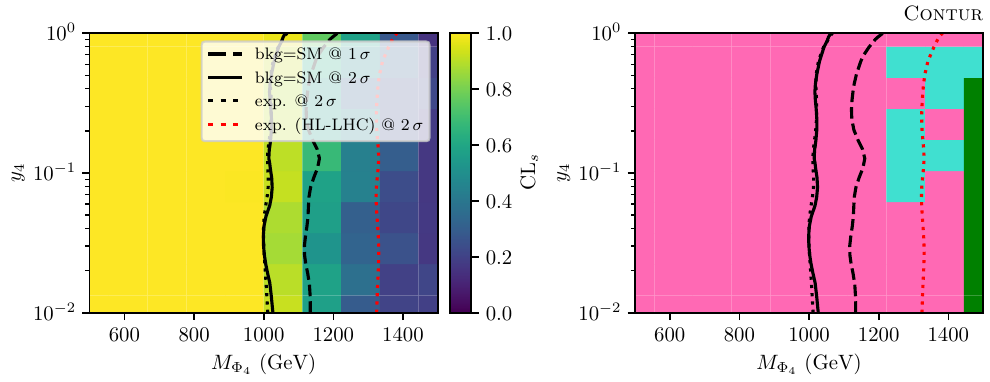}
    \caption{ \label{fig:m4_y4_y2is1_mNis500}}
  \end{subfigure}
  \caption{\contur exclusion plot of $M_{\Phi_4}$ against $y_4$ for 
    (a) $y_2=0$, and (b) $y_2 = 1$.}  
  \begin{tabular}{llll} 
    \swatch{green}~\MET{}+jet  & 
    \swatch{turquoise}~$\ell_1\ell_2$+\MET{}+jet \cite{ATLAS:2024aht} & 
    \swatch{hotpink}~$\tau^+\tau^-$ \cite{ATLAS:2025oiy} \\ 
  \end{tabular}
\end{figure}

Fig.~\ref{fig:m4_y4_y2is0_mNis500} shows a scan of the sensitivity as a function of $y_4$ and $M_{\Phi_4}$ for $y_2 = 0$. Masses
below about 1150~GeV are excluded, with the sensitivity increasing slowly with $y_4$. The most sensitive measurement
is the di-$\tau$ cross section over the whole parameter plane.
Fig.~\ref{fig:m4_y4_y2is1_mNis500} shows the same scan for $y_2 = 1$.
The sensitivity is somewhat reduced, since as seen in Eq.~(\ref{eq:phi23Pdecay}), some of the $\phi^{2/3}_4$ LQs now
decay to tops rather than $\tau$-leptons. Even so, masses up to about 1100~GeV are excluded, with the sensitivity again
increasing slowly with $y_4$. The \MET and top measurements play a slightly more significant role compared to the $y_2 = 0$ case.
\subsubsection{Limits and sensitivity when both $\Phi_3$ and $\Phi_4$ are accessible}
\begin{figure}[t]
  \centering
    \begin{subfigure}[t]{\linewidth}  
      \includegraphics[width=0.80\linewidth]{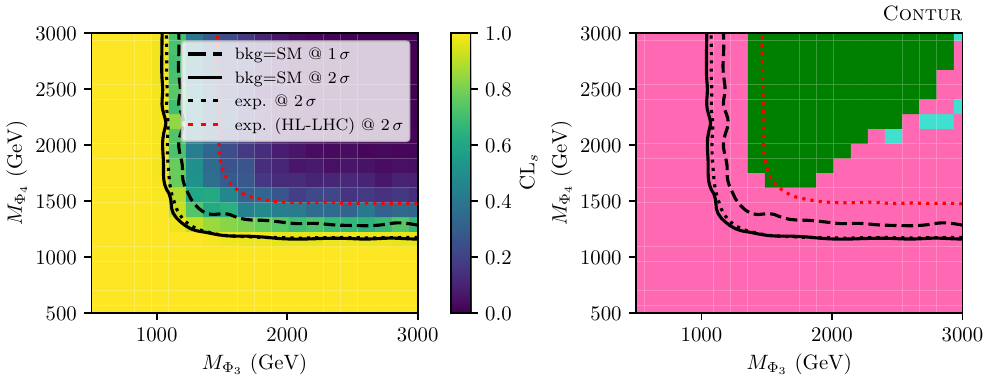}
      \caption{\label{fig:m3_m4_y4is1_y2is0}}
    \end{subfigure}
    \begin{subfigure}[t]{\linewidth}  
      \includegraphics[width=0.80\linewidth]{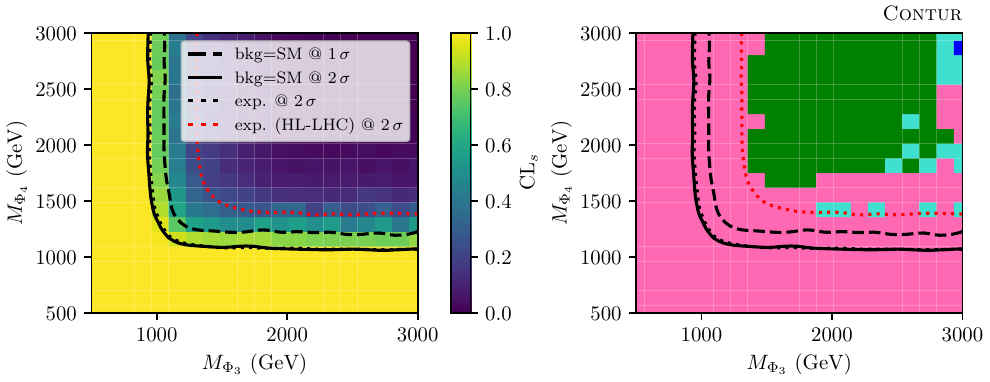}
      \caption{\label{fig:m3_m4_y4is1_y2is1}}
    \end{subfigure}
  \caption{\contur exclusion plot of $M_{\Phi_3}$ against $M_{\Phi_4}$ at $y_4=1$, $m_N = 500$ GeV
  and (a) $y_2=0$ GeV (b) $y_2=1$.}
  \begin{tabular}{llll}
    \swatch{green}~\MET{}+jet \cite{ATLAS:2024vqf} & 
    \swatch{hotpink}~$\tau^+\tau^-$ \cite{ATLAS:2025oiy} & 
    \swatch{turquoise}~$\ell_1\ell_2$+\MET{}+jet \cite{ATLAS:2024aht} \\ 
  \end{tabular}
  \label{fig:m3_m4}
\end{figure}

Fig.~\ref{fig:m3_m4} shows the sensitivity in the plane of $M_{\Phi_3}$ vs $M_{\Phi_4}$, for $M_N = 500$~GeV and
 and {$y_2=0, y_4=1$}, Fig.~\ref{fig:m3_m4_y4is1_y2is0}, and 
$y_2=y_4=1$, Fig.~\ref{fig:m3_m4_y4is1_y2is1}.
 While the phenomenology is more complex in this case, we still see that the same signatures -- $\tau$, top and \MET -- play
 the major role. The increased cross sections mean that when both LQs are around the same mass, the combined reach
 is greater than the individual reaches in the previous section.
\subsubsection{Individual LQ limits}
For completeness, and for comparison with existing searches, scenarios were studied in which the mass hierarchy within a LQ doublet is strong,
such that only one LQ is active at a time.
For $\phi^{-1/3}_3$ alone, and considering the case $y_2=0$, masses below
about 700~GeV are excluded at the 95\% c.l., mainly due to missing transverse momentum signatures.
For $y_2 = 1$ and small $y_4$, there is very little sensitivity due to the
dilution of the \MET signal in $N$ decays, unless $M_N \geq M_{\phi^{-1/3}_3}$.

When $y_2 = 0$, $\phi^{-2/3}_3$ LQ masses below about 1~TeV are excluded at 95\% c.l. by the di-$\tau$ measurement. When $y_2$ is non-zero, the decay channel to a top-quark and heavy neutrino $N$ potentially opens up and the sensitivity is
again strongly reduced unless $M_N \geq M_{\phi^{-2/3}_3}$.
At high $M_{\phi^{-2/3}_3}$, the ATLAS measurement of inclusive four-lepton production~\cite{ATLAS:2021kog} provides some sensitivity
arising from events where $\phi^{-2/3}_3 \rightarrow  \bar{t}N$ and $N \rightarrow t\bar{b}\tau$,
where some of the tops and $\tau$-leptons decay leptonically.

When $y_2 = 0$, the sensitivity to $\phi^{2/3}_4$ is identical to that to $\phi^{-2/3}_3$,
since the only decay open is to $b$-quarks and $\tau$-leptons.
For $y_2 \neq 0$, the decay to SM neutrinos and top quarks opens up and the sensitivity of the $\tau$
measurement drops, but some sensitivity is recovered from top measurements in the dilepton and semileptonic decay channels
as well as from the inclusive $\MET$ measurement. Masses below about 700~GeV are excluded. The only decay open to $\phi^{5/3}_4$ is to top quarks and $\tau$-leptons. Measurements of both provide sensitivity, but
the ATLAS $\tau$ measurement dominates, excluding masses below 1~TeV.
\subsubsection{Existing exclusion limits}
The ATLAS and CMS collaborations have performed dedicated searches for LQ signatures. The QCD  Pair production searches,
which are mostly model independent, provide the most stringent constraints on the LQ mass.
For the 3rd generation up-type scalar LQ ($LQ^u_3$), decaying to $ b\tau$ with $100\%$ branching ratio,
the ATLAS collaboration excludes masses up to 1460 GeV~\cite{ATLAS:2023uox}.
The ATLAS collaboration has also set lower limits on masses up to 1260 GeV~\cite{ATLAS:2021yij} and 1430 GeV~\cite{ATLAS:2021oiz} for
third-generation down-type LQs ($LQ^d_3$) decaying exclusively into $b \nu$ and $t \tau$ final states, respectively.

These bounds on the LQ masses are stronger than the bounds found on individual LQs in this study, which is to be expected since those
dedicated searches focus on specific final-state signatures, and apply multivariate or
machine-learning techniques to enhance signal sensitivity, whereas the bounds derived in \contur~3.1 are from inclusive
cross-section measurements obtained from more general event selections. However, the searches operate on specific benchmark scenarios
and cannot easily be reinterpreted in the more complex phenomenology of this theory, including the presence of
more than one LQ simultaneously, or the opening of decay channels involving the heavy neutrino.

\section{SUMMARY}
\label{summary}
The minimal theory that unifies quarks and leptons at the low energy scale has been investigated.
This theory makes testable predictions at the LHC and future colliders, introducing several new fields beyond those of the Standard Model:
one vector leptoquark, two scalar leptoquarks, a color-octet scalar, and an additional Higgs doublet.
Neutrino masses are generated through the inverse seesaw mechanism,
which naturally keeps them light while allowing the quark-lepton unification scale to remain close to the TeV scale; as a consequence the heavy neutrinos couple to some of the leptoquark states.
The new leptoquarks establish the connection between the quark and lepton sectors and can give rise to distinctive collider
signatures.

In this study we have focussed primarily on the phenomenology of the scalar leptoquarks, which could be produced at energies accessible
to the LHC.
Their dominant decay channels are predicted to involve third-generation particles -- bottom and top quarks, tau leptons, and neutrinos,
sometimes produced directly, and sometimes via a decay chain involving the heavy neutrino.
We computed the production cross sections and branching ratios for these states, and examined how current and future LHC data constrain
the relevant parameter space.
Detailed collider simulations were performed to compare theoretical predictions with existing measurements, with the dominant
sensitivity coming from final states containing third-generation fermions and/or missing transverse momentum.
We find that current LHC measurements and searches place strong lower bounds on the leptoquarks masses in some scenarios,
but that for some reasonable choices of model parameters this sensitivity is drastically reduced, especially when leptoquark
decays proceed via the heavy neutrino.
It remains possible that evidence for, or discovery of, this theory could be obtained in analysis of data from upcoming collider runs. 
In particular, the HL-LHC could probe a large fraction of the parameter space and we could discover this theory for quark-lepton unification in the near future.

{\small {\textit{Acknowledgments:}}
P.F.P. thanks the theory group at Caltech for hospitality at the end of this project. H.D. thanks B.
  Fuks for answering questions about FeynRules while implementing this model in FeynRules.
  This work made use of the High Performance Computing Resource in the Core Facility for Advanced Research
  Computing at Case Western Reserve University and the HEP compute cluster at UCL.
  P.W. thanks STFC for a doctoral training grant.

\bibliography{refs}

\end{document}